\documentclass[fleqn,10pt,a4paper]{wlscirep}

\usepackage[utf8]{inputenc}
\usepackage[T1]{fontenc}
\usepackage{lineno}

\usepackage{threeparttable, tablefootnote}
\usepackage{setspace}
\usepackage{upgreek}
\usepackage{fancyvrb}
\usepackage{xcolor}

\definecolor{keyclr}{rgb}{0.000000, 0.000000, 0.635294}
\definecolor{stringclr}{rgb}{0.558215, 0.000000, 0.135316}

\newcommand{\gb}[1]{\textcolor{green!50!black}{#1}}
\newcommand{\ky}[1]{\textcolor{keyclr}{#1}}
\newcommand{\sg}[1]{\textcolor{stringclr}{#1}}

\usepackage{pinlabel}

\usepackage{color, soul}  
  
\title{Neonatal EEG graded for severity of background abnormalities in hypoxic-ischaemic
  encephalopathy}

\author[1,2,*]{John M O'Toole}
\author[1,2]{Sean R Mathieson}
\author[1,2]{Sumit A Raurale}
\author[1,2]{Fabio Magarelli}
\author[1,3]{William P Marnane}
\author[1,3]{Gordon Lightbody}
\author[1,2]{Geraldine B Boylan}

\affil[1]{INFANT Research Centre, University College Cork, Ireland}
\affil[2]{Department of Paediatrics and Child Health, University College Cork, Ireland}
\affil[3]{Department of Electronic and Electrical Engineering, University College Cork, Ireland}

\affil[*]{corresponding author: JM O'Toole (JOToole@ucc.ie)}

\begin{abstract}
  This report describes a set of neonatal electroencephalogram (EEG) recordings graded
  according to the severity of abnormalities in the background pattern.  
  The dataset consists of 169 hours of multichannel EEG from 53 neonates recorded in a
  neonatal intensive care unit.
  All neonates received a diagnosis of hypoxic-ischaemic encephalopathy (HIE), the most
  common cause of brain injury in full term infants.  
  For each neonate, multiple 1-hour epochs of good quality EEG were selected and then
  graded for background abnormalities.
  The grading system assesses EEG attributes such as amplitude and frequency, continuity,
  sleep--wake cycling, symmetry and synchrony, and abnormal waveforms.
  Background severity was then categorised into 4 grades: normal or mildly abnormal EEG,
  moderately abnormal EEG, severely abnormal EEG, and inactive EEG.
  The data can be used as a reference set of multi-channel EEG for neonates with HIE, for
  EEG training purposes, or for developing and evaluating automated grading algorithms.
\end{abstract}
\begin{document}


\flushbottom
\maketitle

\onehalfspacing
\thispagestyle{empty}

\section*{Background \& Summary}
Impaired oxygen delivery or blood flow to the brain around the time of birth can cause
brain injury.
Infants develop an encephalopathy called hypoxic-ischaemic encephalopathy (HIE), which is
the leading cause of death and disability in full term neonates.
Incidence rates of HIE are around 2 per 1,000 deliveries in high-income countries with
higher rates in low- to middle-income countries \cite{Kurinczuk2010}.
HIE can cause neonatal death or significant neurological and neurodevelopmental impairment
such as cerebral palsy, epilepsy, or learning disabilities \cite{Perez2013}.
HIE is an evolving brain injury.
The primary injury is followed by a latent phase which lasts for approximately 6 hours. 
This is followed by the secondary injury phase, a delayed phase of programmed
cell death.
Therapeutic hypothermia is the only intervention available for infants with moderate to
severe HIE and it must be instigated before the onset of the secondary phase of injury if
it is to be effective. 

The electroencephalogram (EEG) allows for continuous cot-side monitoring of cerebral
function.
A hypoxic-ischaemic insult can alter the normal background pattern of the EEG, providing a
unique insight into cerebral dysfunction \cite{Walsh2011}.
This deviation from normal EEG background is associated with adverse neurodevelopmental
outcome \cite{Holmes1982,Pressler2001,Murray2009a}.
As EEG is a valuable measure of severity of ongoing encephalopathy, it can be
particularly beneficial when commenced within the primary phase of injury to help determine
which infants may benefit from therapeutic hypothermia \cite{Tagin2012}.

Review of the EEG requires specialist expertise not always available in neonatal intensive
care units.
Computer-based methods have the potential to automate the process of grading background
EEG activity for severity of injury.
These automated methods could produce a continuous objective measure of EEG activity that
could be easily scaled to monitor a high-volume of neonates, far beyond what would be
humanly possible.
Many methods have been developed to generate background grading systems
\cite{Korotchikova2011,Ahmed2016,Stevenson2013,Raurale2021a,Raurale2019,Matic2014,Matic2015,Guo2020,Raurale2021,Moghadam2021}.
This existing body of work highlights the potential of signal processing and machine
learning methods to construct accurate classifiers of background EEG.
Despite this significant progress, more can be achieved in this area.
Thus far, progress has been confined to individual research groups pursuing different
approaches.
Comparing methods is difficult for many reasons \cite{Moghadam2021}, including the lack of
an accepted standard grading scheme \cite{Walsh2011} and freely-available EEG data.
Aiming to address some of these limitations---and inspired by the success of an open-access
neonatal EEG data set with annotations \cite{Stevenson2019}---we present an open-access
EEG data set recorded within the first days after birth for infants with a HIE diagnosis.
Multiple 1-hour EEG epochs for each infant were graded for severity of background
abnormality.
This data could be used to develop new algorithms or benchmark existing ones.
The data could also be used to assist in training of the review of background neonatal
EEG.

\section*{Methods}

\subsection*{Patients}

A subset of EEG records were retrieved from data collected during a medical-device trial. 
The clinical investigation evaluated the effectiveness of a machine-learning algorithm to
detect seizures \cite{Pavel2020,Pavel2021}. 
Neonates that were clinically determined to be at-risk of seizures, with a gestational age
between 36 to 44 weeks, and admitted to the neonatal intensive care unit (NICU) were
considered for inclusion in the study. 
After written and informed consent from a guardian or parent, neonates were enrolled over
a period from Januray 2011 to Feburary 2017. 
Data were collected across 8 neonatal centres in Ireland, the Netherlands, Sweden, and the
UK.

As part of the medical-device trial, 472 neonates were recruited \cite{Pavel2021}.
From this group, 284 neonates were selected with a clinical HIE diagnosis and a valid EEG
recording of at least 6 hours in duration.
Eighteen infants were excluded because of a combined diagnosis, 68 were held out for a
future validation set for development of an EEG algorithm, and 17 were excluded because
the EEG did not start within 48 hours after birth.
From the remaining 181, we only included 54 neonates for which we had permission to share
the data. 
That is, the EEGs that were recorded in Cork University Maternity Hospital, Ireland.
After closer examination of the EEG, a further neonate was excluded due to the low-quality
of the EEG recording.
For this cohort of 53 neonates, the median gestational age was 40 weeks, most were male
(62\%), and most (59\%) received therapeutic hypothermia, as presented in
Table~\ref{tab:clin_char}.  

The study to collect EEG data at Cork University Maternity Hospital was approved by the
Cork Ethics Research Committee. 
The same ethics committee also approved the Open Access release of the fully and
irrevocably anonymised EEG recordings. 
Permission to share the data was obtained from the Data Protection Officer at
University College Cork, Ireland.

\begin{table}[ht]
  \centering
  \begin{threeparttable}    
  \begin{tabular}{ll}
    \toprule
                                             & $n=53$                 \\
    \midrule
    Gestational age (weeks)                  & 40.0 (39.4 to 40.7)    \\
    Birth weight (g)                         & 3,470 (3,190 to 3,800) \\
    Sex (male)                               & 33 (62\%)              \\  
    Sarnat score at 24 hours:\tnote{$\dagger$} &                        \\
    \hspace{1em} mild                        & 23 (43\%)              \\
    \hspace{1em} moderate                    & 18 (34\%)              \\
    \hspace{1em} severe                      & 8 (15\%)               \\
    Therapeutic hypothermia:                 &                        \\
    \hspace{1em} cooled                      & 31 (59\%)              \\
    \bottomrule 
  \end{tabular}
  \begin{tablenotes}\footnotesize
  \item Data represented as median (interquartile range) or number (\%).
    \item[$\dagger$] $n=49$
    \end{tablenotes}
  \end{threeparttable}
  \caption{Clinical characteristics}  
  \label{tab:clin_char}
\end{table}

\subsection*{EEG}

EEG was recorded as soon as possible after birth for a prolonged period up to 100 hours
after birth.
Two EEG machines were used, the NicoletOne ICU Monitor (Natus, Middletion, WI, USA) for 24
neonates and the Neurofax EEG-1200 (Nihon Khoden, Tokyo, Japan) for 29 neonates.
EEG was sampled at a rate of 256 Hz (NicoletOne) and 200 Hz (Neurofax).
Disposal electrodes were placed over the central (C3 and C4), frontal (F3 and F4),
occipital (O1 and O2), and temporal (T3 and T4) regions and at the midline (Cz), using a
reduced version of the 10:20 international system \cite{Pavel2021}.
EEG was recorded relative to a reference channel: an average between C3 and C4 for the
Neurofax recordings and FCz, a mid-line placement between Fz and Cz, for the NicoletOne
recordings.

\begin{figure}[ht]
  \centering
  \includegraphics[scale=1.0]{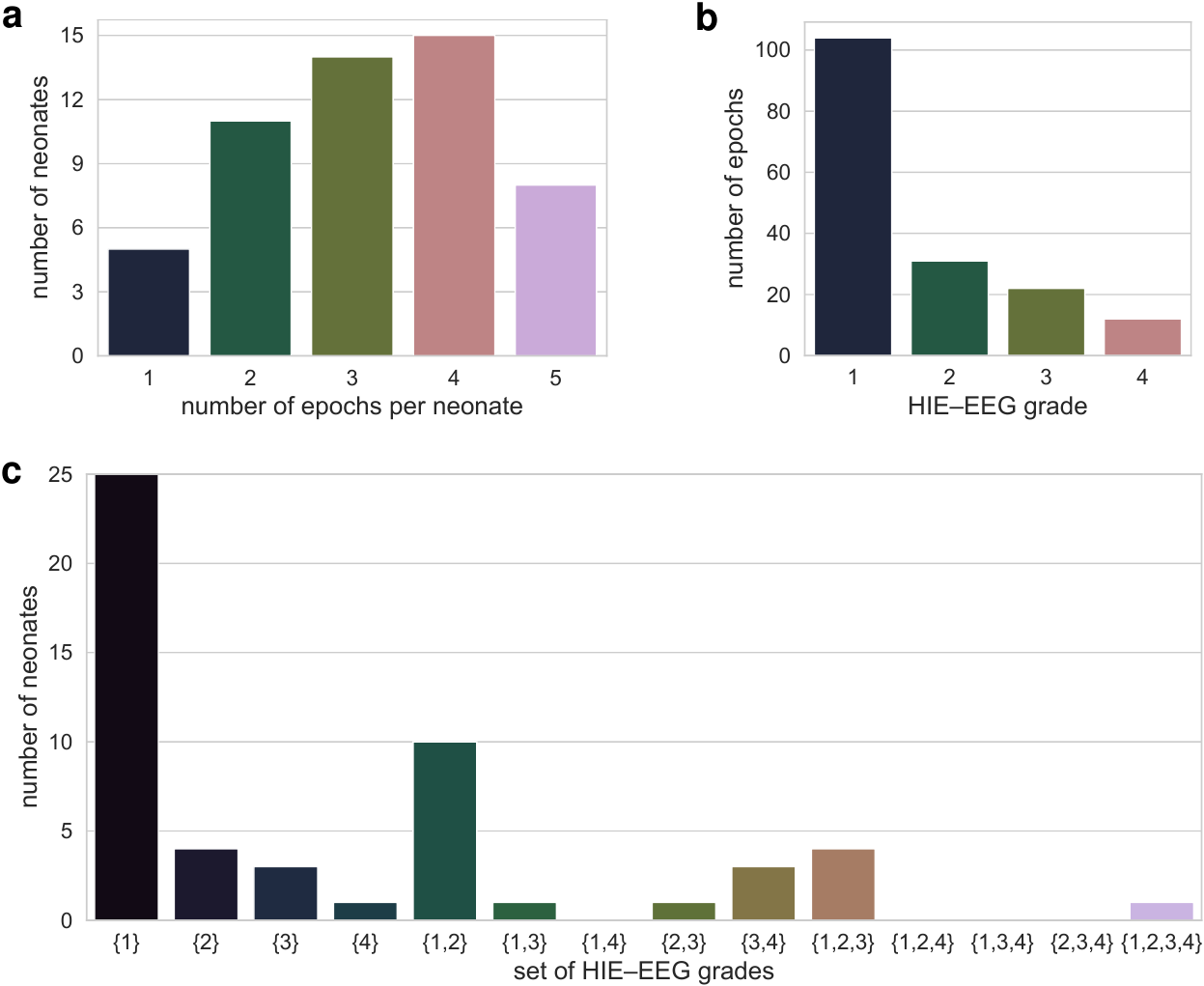}
  \caption{Distribution of EEG epochs. One-hundred and seventy one 1-hour epochs were
    pruned from continuous EEG recordings from 53 neonates. 
    Distribution of epochs per baby in ({\bf a}) and grades of EEG hypoxic-ischaemic
    encephalopathy (HIE) in ({\bf b}). Some neonates have more than one grade throughout the
    epochs: ({\bf c}) illustrates the distribution of all possible combinations of sets of grades per neonate.}
  \label{fig:eeg_epochs}
\end{figure}

EEG was exported from the proprietary format of the NicoletOne and Neurofax machines to
the open European Data Files (EDF) format and securely stored for off-line analysis.
All data was fully anonymised. 
For each neonate, a maximum of 5 1-hour epochs were pruned from the continuous EEG
recording.
Artefacts are not uncommon in long-duration EEG recorded in a busy intensive care
environment.
Epochs were selected to avoid as much artefact as possible and for all epochs the majority
of the epoch was artefact free.
They were distributed in time throughout the duration of continuous recording but limited
to the first 48 hours after birth.
In total 169 epochs, exactly 60 minutes in duration, were included in the data set.
The median number of epochs per neonate was 3, with an inter-quartile range of 2 to 4.
Fig.~\ref{fig:eeg_epochs}(a) illustrates the distribution of epochs per neonate.

\begin{figure}[ht]
  \centering
  \includegraphics[width=\textwidth]{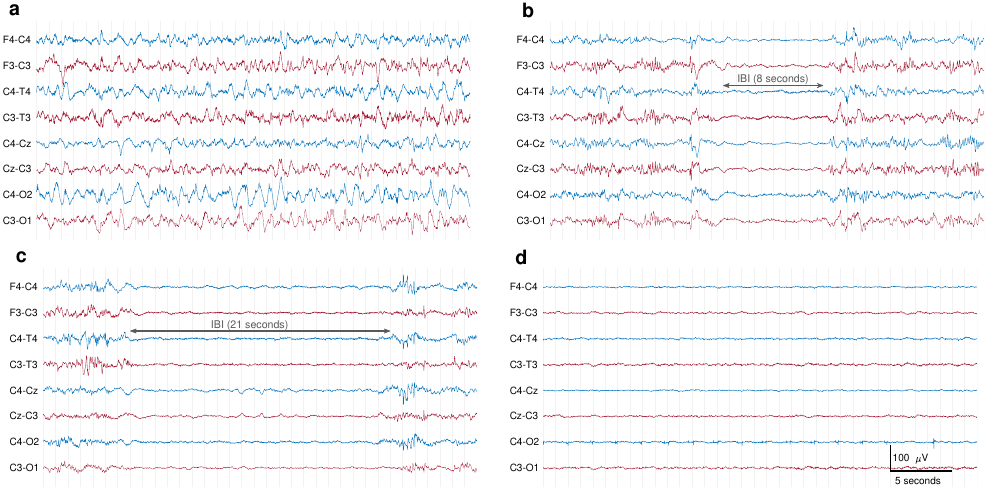}
  \caption{Examples of different EEG grades. Thirty-six seconds of EEG from different
    neonates. ({\bf a})
    normal or mildly abnormal EEG (grade 1); ({\bf b}) moderately abnormal EEG (grade 2);
    ({\bf c}) major abnormalities (grade 3); ({\bf d}) inactive EEG (grade 4). Inter-burst
    intervals (IBI) are annotated in the grade 2 and grade 3 examples. All EEGs are in bipolar
    montage, plotted with the same time and amplitude scale, and bandpass filtered from
    0.3 to 35 Hz.}
  \label{fig:eeg_grades}
\end{figure}

Two clinical physiologists with expertise in neonatal EEG (authors SRM and GBB)
independently graded each epoch using a commonly-used EEG classification scheme
\cite{Murray2009a, Korotchikova2011, Ahmed2016, Raurale2021}.
Table~\ref{tab:eeg_grades} provides an outline of this grading system.
Where grades disagreed between the 2 experts, they jointly reviewed the epoch and decided
on a consensus grade.
The grading system includes measures of varying degrees of discontinuous activity, normal
and abnormal patterns, symmetry and synchrony of activity across the hemispheres, and the
quality or lack of sleep--wake cycling.
Full details of the grading system can be found in Murray \emph{et al.} \cite{Murray2009a}.
Although seizures are not part of the grading system, some epochs did contain
short-duration seizures. 
Due to the short-duration nature of these seizures comparative to the 1-hour epoch, there
was sufficient background activity to assign a grade that was not solely based on the
presence of seizures.  
Grade 0 (normal EEG) and grade 1 (mildly abnormal EEG) were combined into a new grade 1, to represent both
normal and mildly abnormal EEG.
The distribution of these 4 grades for the 169 epochs is illustrated in
Fig.~\ref{fig:eeg_epochs}(b).
The vast majority of epochs are grade 1: 104 for grade 1, 31 for grade 2, 22 for grade 3,
and 12 for grade 4.
Not all neonates had the same grade throughout all epochs, as illustrated in
Fig.~\ref{fig:eeg_epochs}(c).
The most common set of grades was $\{1\}$ ($n=25$), followed by $\{1,2\}$ ($n=10$) and
$\{1,2,3\}$ ($n=4$). One neonate had all 4 grades across 5 epochs.
Example EEG segments for each grade are presented in Fig.~\ref{fig:eeg_grades}.

\begin{table}[ht]
  \centering
  \begin{threeparttable}
  \begin{tabular}{lllll}
    \toprule
    EEG grade (description)        & Background           & IBI (s) & Features of the EEG
                                   & Sleep--wake cycle                                                                                        \\
    \midrule
    0     (normal)                 & Continuous           & --      & Normal physiologic features                       &                     \\
                                   &                      &         & (e.g. anterior slow waves)                        &                     \\
    1     (mild abnormalities)     & Continuous           & --      & Slightly abnormal activity                        & Or poorly defined   \\
                                   &                      &         & (e.g. mild asymmetry,                             &                     \\
                                   &                      &         & mild voltage depression)                          &                     \\
    2     (moderate abnormalities) & Discontinuous        & $<$10   & Or clear asymmetry                                & Not clearly defined \\
                                   &                      &         & or asynchrony                                     &                     \\
    3     (major abnormalities)    & Discontinuous        & 10--60  & Severe attenuation of                             & Or absent           \\
                                   &                      &         & background patterns                               &                     \\
    4     (inactive)               & Severe discontinuity & $>$60   & Or background activity of $<$10 $\upmu\mathrm{V}$ & Absent              \\
    \bottomrule
  \end{tabular}
  \begin{tablenotes}
    \footnotesize
    \item IBI: inter-burst interval.     
    \end{tablenotes}
  \end{threeparttable}
  \caption{EEG Classification Adapted from Murray \emph{et al.} \cite{Murray2009a}}  
  \label{tab:eeg_grades}
\end{table}

\section*{Data Records}

The EEG data with grades for each epoch is available at Zenodo
(\url{https://doi.org/10.5281/zenodo.6587973}) \cite{o_toole_john_m_2022_6587973}.
Data are provided as European data format (EDF) files and as compressed comma separated
values (CSV) files.
The EDF files are stored in the \Verb|EDF_format/| folder and the CSV files are stored in
the \Verb|CSV_format/| folder. 
Each 1-hour epoch is stored as a separate file, using the file name convention
\Verb|IDXX_epochY|.
For example, file \Verb|ID10_epoch3| is the 3rd epoch for baby 10.
A separate file called \Verb|eeg_grades.csv|, in CSV format, contains the grades assigned
to each epoch.
Another CSV file (\Verb|metadata.csv|) contains additional information on the epochs: a
description of the quality of the EEG, whether seizures are present or not, reference
electrode used in the recording, and the sampling frequency.

\section*{Technical Validation}

\begin{figure}[ht]
  \centering
  \includegraphics[width=\textwidth]{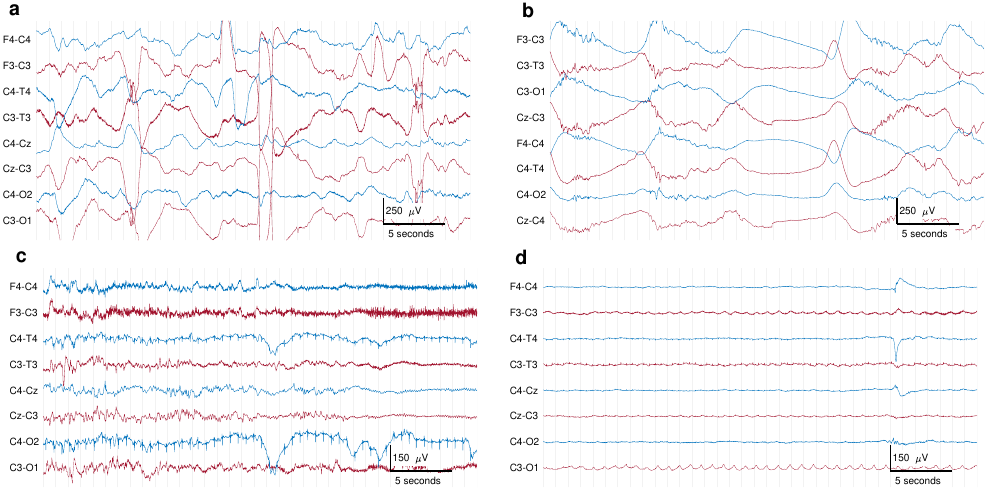}
  \caption{Examples of some typical EEG artefacts. 
    Thirty-six seconds of EEG from different neonates. EEG segment with ({\bf a})
    high-amplitude movement artefact (grade 1 EEG); ({\bf b}) sweat artefact (also grade 1
    EEG);({\bf c}) ECG artefact on C4--O2 and C4--T4 and high-frequency
    artefact, most prominent on F3--C3 (grade 3 EEG); ({\bf d}) respiration artefact
    across the left hemisphere channels, and clearly visible on C3--O1 (grade 4 EEG).
    All EEGs are in bipolar montage and bandpass filtered from 0.1 to 35 Hz.}
  \label{fig:eeg_artefacts}
\end{figure}

EEG was recorded according to clinical standards in the NICU.
The epochs of EEG were selected to be of high quality, with reduced artefact.
However, as these EEGs are recorded in a busy intensive-care environment, they are not
completely free from artefact.
Fig.~\ref{fig:eeg_artefacts} shows an example of some artefacts.
The types of artefacts vary from biological origin, such as sweat artefacts, muscle, or
respiration; to artefacts of non-biological origin, such as 50 Hz power-supply
interference from nearby devices.  
Periodic checks of the electrode impedance causes a pause of EEG recording, resulting in
periods of flat, near-zero EEG.
The median duration of flat trace in the 60-minute epochs was 0.7 seconds, with an
interquartile range of 0.4 to 1.0 seconds and a range of 0 to 431.1 seconds.

  As part of a quality check of the EEG recording, we compare EEG power at lower and
  higher frequencies. The lower-frequency activity is a measure of cortical activity
  whereas the higher-frequency is unlikely to measure cortical activiy and more likely to
  be nothing more than the noise floor \cite{Stevenson2019}.
  We calculate power per channel across the bipolar montage F3--C3, T3--C3, O1--C3,
  C3--Cz, Cz--C4, F4--C4, T4--C4, and O2--C4. 
  Each channel is bandpass filtered with an infinite-impulse response (IIR) filter, a type
  II Chebyshev filter of order 21. 
  Power is then calculated within the passbands 0.5 to 16 Hz (low frequency) and 77 to 99
  Hz (high frequency).
  The median power, over all epochs, for the low-frequency band is 175.4
  $\upmu\mathrm{V}^2$ (interquartile range, IQR: 83.7 to 419.8 $\upmu\mathrm{V}^2$) and
  8.4 $\upmu\mathrm{V}^2$ (IQR: 1.6 to 36.2 $\upmu\mathrm{V}^2$) for the higher frequency
  band.
  Thus we find that our estimates of the noise floor is considerable lower than the
  recording of EEG cerebral activity.

\begin{figure}[ht]
  \centering
  \includegraphics[scale=1]{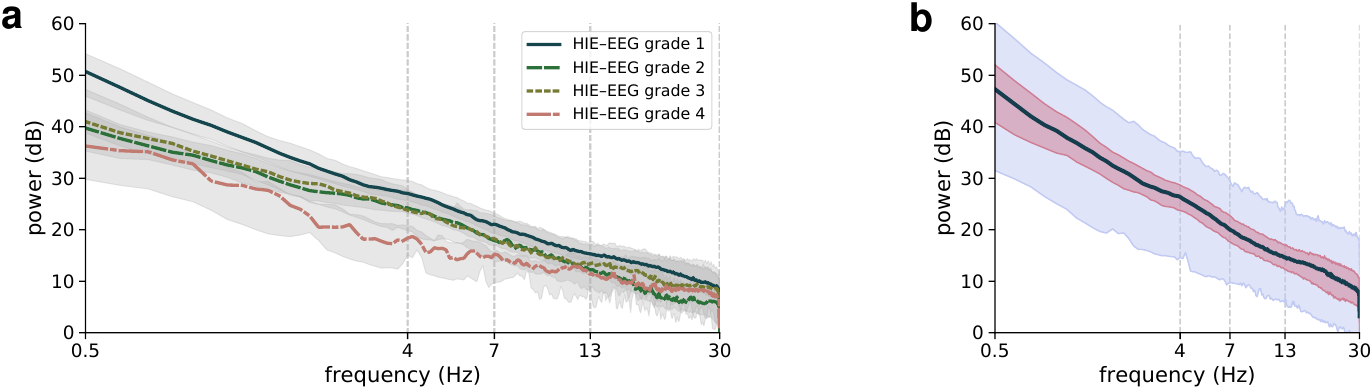}
  \caption{Spectra plotted on a log--log scale from epochs for each grade ({\bf a}) and a
    grand-average for all epochs ({\bf b}). 
    Thick lines represent the median value across all epochs, and shaded areas represent
    the inter-quartile range in ({\bf a}) and ({\bf b}) and 95-th centile range in ({\bf b}).
    There are $104$ epochs for grade 1, $31$ for grade 2, $22$ for grade 3, and $12$ for
    grade 4. 
  }
  \label{fig:eeg_psds}
\end{figure}

Next, to further validate the technical quality of the EEG, we computed the frequency
response for all epochs and generated a set of quantitative EEG (qEEG) features
\cite{Finn2018,OToole2017b,OToole2019a}.
Power spectral densities (PSD) were estimated using the Welch method with an 8-second
Hamming window and 75\% overlap.
PSDs were calculated per channel using the same bipolar montage described previously.
Each channel's estimate is then averaged over all 8 channels for the 1-hour epoch.
Fig.~\ref{fig:eeg_psds} summarises the PSDs for all epochs per grade.
Grades 1 to 3 indicate a linear log--log frequency response, known as a power-law
response, in keeping with current understanding of neonatal EEG
\cite{Finn2018,Korotchikova2009,Stevenson2010b,Stevenson2013}.
For grade 4, the response appears more nonlinear, but the lower number of epochs in this
group (12, compared with 23, 31, and 105) may be a factor here.

The qEEG feature set consisted of 5 features: spectral power, range-EEG (rEEG),
interhemispheric coherence, fractal dimension, and spectral edge frequency
\cite{OToole2017b,Finn2018,OToole2019a}.
Features are estimated using the same bipolar montage described previously.
Spectral power and coherence features are generated separately in 4 frequency bands: delta
(0.5--4 Hz), theta (4--7 Hz), alpha (7--13 Hz), and beta (13--30 Hz).
The rEEG is calculated within the 1--20 Hz bandwidth and assessed at the lower-, median-,
and upper-margins \cite{OToole2017b}.
Inter-hemispheric coherence is an averaged value of coherence calculated between the
following channel pairs: F3--C3 and F4--C4, T3--C3 and T4--C4, and O1--C3 and O2--C4.
Spectral power, coherence, fractal dimension, and spectral edge frequency (95\%) are
estimated on a 64 second segment of EEG with 50\% overlap.
The median value of all segments is used to summarise the feature over the 1-hour epoch.
All features, excluding coherence, are estimated on a channel by channel basis and
summarised by the median value across channels.
Features were generated using the NEURAL toolbox (version 0.4.4) \cite{OToole2017b}.  

\begin{figure}[ht]
  \centering
  \includegraphics[width=\textwidth]{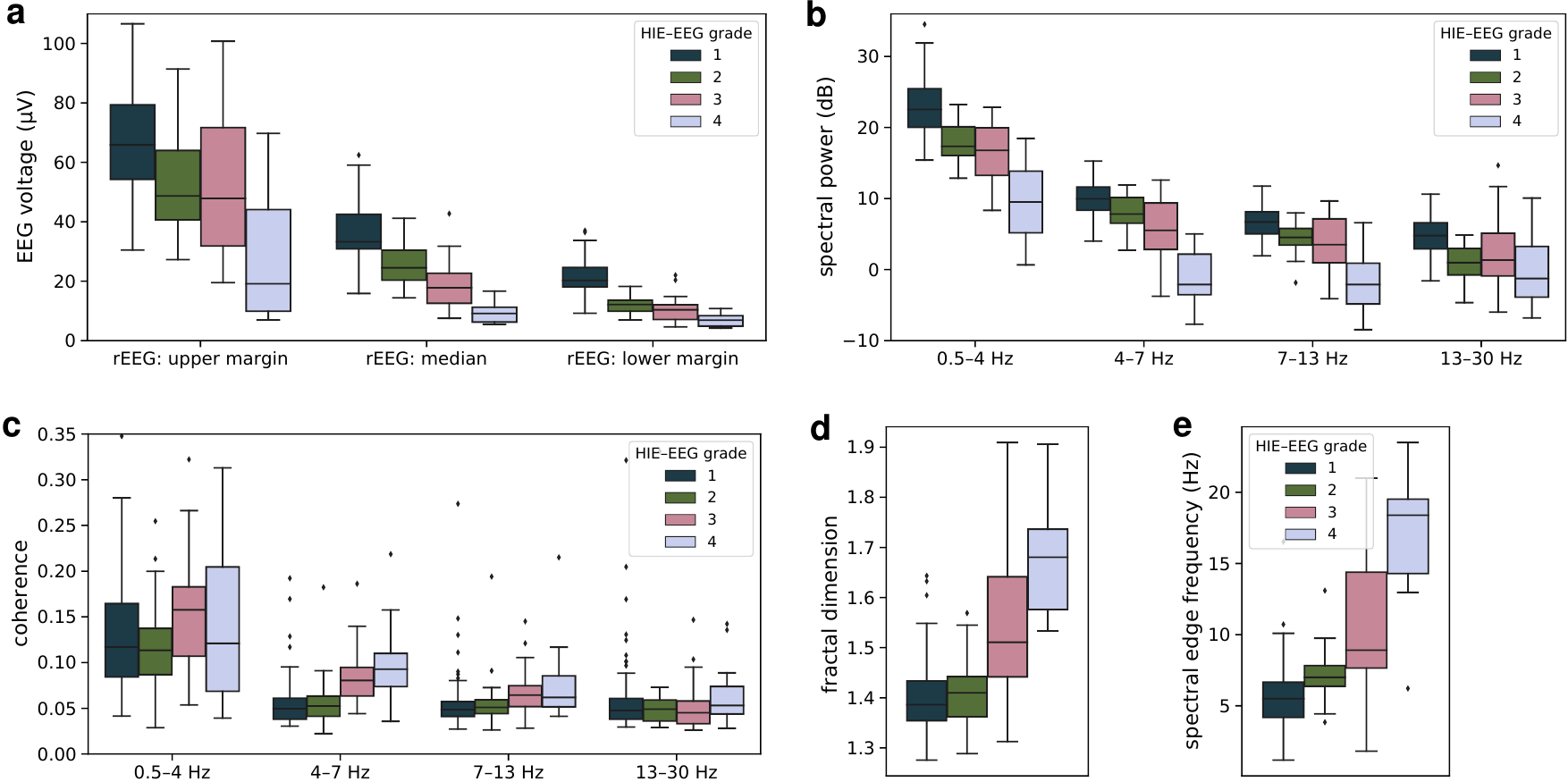}
  \caption{Quantitative summary measures of the EEG for the 4 HIE--EEG grades. 
    Three summary measures of the range-EEG (rEEG) in ({\bf a}), spectral power in ({\bf
      b}), inter-hemispheric coherence in ({\bf c}), fractal dimension in ({\bf d}), and
    spectral edge frequency in ({\bf e}). 
    Spectral power and coherence features are calculated for 4 different frequency bands.}
  \label{fig:eeg_qEEG}
\end{figure}

Fig.~\ref{fig:eeg_qEEG} plots the distribution of the 5 features, highlighting the
differences for many features across the 4 grades.
The rEEG in Fig.~\ref{fig:eeg_qEEG}(a), a measure of peak-to-peak voltage, shows
decreasing EEG amplitude through the 4 grades, with the difference between the grades
particularly pronounced in the median rEEG feature.
Similarly, the difference in spectral power for the 4 grades across the 4 frequency bands
is evident in Fig.~\ref{fig:eeg_qEEG}(b).
Significant, although low-valued, interhemispheric coherence is present across the 4
frequency bands, with higher levels of coherence in the delta (0.5--4 Hz) band compared to
the other 3 frequency bands.
Both the fractal dimension and spectral edge frequency plots indicate a difference in
spectral shape for the grades, in the form of a decreasing slope of the log--log spectra
with increasing grades.

\section*{Usage Notes}

The EEG files are stored in both EDF and CSV format.  
The EDF format was developed in 1992 for sleep EEG files and has remained a standard
open-format for EEG files \cite{Kemp1992}.
EDF files can be viewed in most EEG review software, including free versions such as
\emph{EDFbrowser} (\url{https://www.teuniz.net/edfbrowser/}).
The format stores data in 16-bit integers and therefore will likely be converted to 64-bit
floating-point numbers before analysis or viewing of the data.
Storing EEG in 16-bit integers reduces the file size comparative to 64-bit floating-point
numbers but does so in a lossy manner; however many lossless compression algorithms can
now match this compression without loss of information.
Despite its outdated structure, it still remains a standard open format for EEG review.

For analysis using software tools, special libraries are required to read EDF files as
this format is not used for other data types.
For this reason, to simplify the process of loading the data for analysis we also provide
the widely accessible CSV format.
Our CSV format stores time (in seconds) and voltage (in micro-Volts) at each of the 9
channels.
The first line of each CSV file contains a header with the name of each column.
These files are compressed using the cross-platform XZ format, which uses the
Lempel--Ziv--Markov chain algorithm.
Freely available compression tools can be used to decompress the files. For example, 7z
(\url{https://www.7-zip.org/}) for Windows operating systems or XZ Utils
(\url{https://tukaani.org/xz/}) for Linux operating systems.
Alternatively, many programming languages can provide this decompression when importing
the data.
In the Python programming environment (Python Software Foundation,
\url{https://www.python.org/}) for example, the \Verb|read_csv| function from the Pandas
package can directly read in the tabular data compressed with XZ \cite{pandas2022}:
\begin{Verbatim}[commandchars=\\\{\}]
\textbf{\gb{import}}\ 
\textbf{\ky{pandas}}\
\textbf{\gb{as}}\
\textbf{\ky{pd}}
eeg_df = \ky{pd}.read_csv(\sg{"ID10_epoch3.csv.xz"})
\end{Verbatim}
for example file \Verb|ID10_epoch3.csv.xz|.
Likewise, the R programming environment (R Core Team, \url{https://www.R-project.org/}),
the \verb|read.csv| function can directly import data in the compressed XZ format.
\begin{Verbatim}[commandchars=\\\{\}]
eeg_df <- read.csv(\sg{"ID10_epoch3.csv.xz"})
\end{Verbatim}
For Matlab (The Mathworks, Inc., United States) and Julia (The Julia Project,
\url{https://julialang.org/}) \cite{bezanson2017julia}, the CSV files must be uncompressed
before importing.
In Matlab,
\begin{Verbatim}[commandchars=\\\{\}]
eeg_tb = readtable(\sg{"ID10_epoch3.csv"});
\end{Verbatim}
where \Verb|ID10_epoch3.csv| is the uncompressed version of \Verb|ID10_epoch3.csv.xz|. In Julia,
\begin{Verbatim}[commandchars=\\\{\}]
\textbf{\gb{using}} \textbf{\ky{CSV}}
\textbf{\gb{using}} \textbf{\ky{DataFrames}}
eeg_df = \gb{CSV}.read(\sg{"ID10_epoch3.csv"}, \ky{DataFrame})
\end{Verbatim}

The data could be used for training purposes. For this, the data can be viewed in a EEG
viewer using both a referential or bipolar montage.
Data are provided in the raw referential format for processing purposes, however visual
analysis is typically conducted using a bipolar montage. The bipolar montage displayed
during recording and used for background scoring contains the following bipolar electrode
pairs: F4--C4, C4--O2, F3--C3, C3--O1, T4--C4, C4--CZ, CZ--C3, C3--T3. This contains both
antero--posterior and transverse elements. Typical display setting for reviewing neonatal
EEG include: sensitivity 70--100 $\upmu\mathrm{V}$/cm, timebase 15--20 mm/sec, and bandpass filter
0.5--70 Hz. Amplitude-integrated EEG (aEEG) channels are popular in clinical review and
within this montage might include aEEG channels for F4--C4, F3--C3 and C4--C3.

The EEG data could also be used to develop an automated EEG grading algorithm.
These classification algorithms use signal processing and machine-learning methods to
extract information from the EEG that is characteristic of the particular grade of EEG
\cite{Korotchikova2011,Ahmed2016,Stevenson2013,Raurale2021a,Raurale2019,Matic2014,Matic2015,Guo2020,Raurale2021,Moghadam2021}.
The first stage in algorithm development is to preprocess the data.
This will include a bandpass filter, typically followed by downsampling.
The bandpass filter, at 0.5 to 30 Hz for example, removes the 50 Hz power-line noise and
very-slow activity (<0.5 Hz) often associated with artefact such as DC drift or sweat
artefact.
The bandpass filter also allows for downsampling without aliasing.
Downsampling is commonly applied to reduce the algorithm's computational load with
negligible loss in performance.
The preprocessed EEG is then ready for use in a classifier to grade the EEG.

There are 2 approaches to developing a grading algorithm which incorporate
machine-learning models.
The first is to use signal-processing methods to extract a set of features from the
preprocessed EEG and then combine these features using a machine-learning model
\cite{Korotchikova2011,Ahmed2016,Stevenson2013,Raurale2021a,Raurale2019,Matic2014,Matic2015,Guo2020,Moghadam2021}.
For this approach, we must develop and curate a set of features that adequately
generalises the main discriminating factors among the 4 grades.
As Fig.~\ref{fig:eeg_qEEG} indicates, there are potentially many different features that could
discriminate, with varying levels of accuracy, between the 4 grades.
The second approach is to use deep-learning methods, which provides an end-to-end (EEG to
grade) solution \cite{Raurale2021, Raurale2020}.
These methods automatically extracts and combines the features in a single neural network.
For this approach, we must the select the type of neural network to use, for example a
convolutional or recurrent neural network, and then design the specific architecture of
the network.

For both approaches, the machine-learning model is constructed using a data-driven
approach through training and testing.
We recommend training and testing the model using some form of cross validation, ideally
leave-one-out.
The split of training and testing data should be done on a neonate level, not on an epoch
level.
This will avoid testing a model that was trained using epochs from the same neonate.
Regardless of which approach is used, there is, unfortunately, no one-size-fits-all model.
That is, different applications will require different models.
Considerable care and attention to the design process is required to develop an accurate
and robust classifier, but certainly worth the effort given the potential clinical utility
of such an algorithm to improve health outcomes for infants with HIE.

\section*{Code availability}

Custom code was not used to generate the data.  
EEG files were exported from proprietary format to EDF files using the associated EEG
reviewing software for the NicoletOne and Neurofax EEG machines. 
Details on how to view the EEG data and import it into programming environments is
described in the Usage Notes section.

To assist with computer-based analysis of the EEG, we provide freely-available code to
downsample the EEG to a lower and uniform sampling rate.
For quantitative or machine-learning analysis, the neonatal EEG is often downsampled to a
lower sampling rate, as the majority of the power is typically below 10 to 20 Hz. 
For example, Fig.~\ref{fig:eeg_qEEG}(e) shows that 95\% of spectral power is below 25Hz.
The processing routines include an anti-aliasing filter before downsampling. 
Both Matlab and Python versions of the code are included at
\url{https://github.com/otoolej/downsample_open_eeg} (commit: 98ba2a1).

\section*{Acknowledgements} 
This work was supported by an Innovator Award from the Wellcome Trust (209325/Z/17/Z). The
medical-device trial which recorded the EEG was supported by a Strategic Translational Award
also from the Wellcome Trust (098983). The article is based upon work by the COST Action
\emph{AI-4-NICU} (CA20214), supported by COST (European Cooperation in Science and
Technology, \url{https://www.cost.eu/}).

\section*{Author contributions statement}
GBB and WPM were leading investigators in the clinical trial which collected the EEG.  
SRM prepared the EEG epochs and GBB and SRM reviewed and graded all epochs.
JMOT conducted the analysis of the EEG.  
JMOT, SRM, and FM prepared and formatted the data for open access.
JMOT produced the first draft and all authors critically reviewed and revised the manuscript.

\section*{Competing interests} 
The authors have no competing interests to declare.


\end{document}